\documentclass[mnsc,nonblindrev]{informs3} 

\OneAndAHalfSpacedXI 
\usepackage{hyperref}

\usepackage{endnotes}
\let\footnote=\endnote
\let\enotesize=\normalsize
\def\notesname{Endnotes}%
\def\makeenmark{$^{\theenmark}$}
\def\enoteformat{\rightskip0pt\leftskip0pt\parindent=1.75em
  \leavevmode\llap{\theenmark.\enskip}}

\usepackage{natbib}
 \bibpunct[, ]{(}{)}{,}{a}{}{,}%
 \def\bibfont{\small}%

\usepackage{algorithm}
\usepackage{algorithmic}
\makeatletter
\usepackage{makecell}

\usepackage{ulem}
\usepackage{booktabs}
\usepackage{multirow}
\usepackage{comment}
\usepackage{graphics}
\usepackage{graphicx}
\usepackage{subfigure}

\usepackage{xcolor}
\hypersetup{
    colorlinks,
    linkcolor={red!50!black},
    citecolor={blue!50!black},
    urlcolor={blue!80!black}
}
\TheoremsNumberedThrough     
\ECRepeatTheorems

\EquationsNumberedThrough    

\newenvironment{assumption'}[1]
  {\renewcommand{\theassumption}{\arabic{assumption}$'$}%
   \addtocounter{assumption}{-1}%
   \begin{assumption}}
  {\end{assumption}}

\newcommand{\indep}{\perp \!\!\! \perp}

\usepackage{amsmath}
\usepackage{amsfonts}
\usepackage{amssymb}
\usepackage{graphics}
\usepackage{graphicx}
\usepackage{dsfont}
\newenvironment{alg'}[1]
  {\renewcommand{\thealgorithm}{\ref{#1}$'$}%
   \addtocounter{algorithm}{-1}%
   \begin{breakablealgorithm}}
  {\end{breakablealgorithm}}

\newenvironment{lemma'}[1]
  {\renewcommand{\thelemma}{\ref{#1}$'$}%
   \addtocounter{lemma}{-1}%
   \begin{lemma}}
  {\end{lemma}}

\begin{document}



\RUNTITLE{RT}

\TITLE{Causal inference with Machine Learning-Based Covariate Representation}


\ARTICLEAUTHORS{%
\AUTHOR{Yuhang Wu, Jinghai He, Zeyu Zheng}
\AFF{Department of Industrial Engineering and Operations Research, University of California, Berkeley}
} 

\ABSTRACT{Utilizing \textit{covariate} information has been a powerful approach to improve the efficiency and accuracy for causal inference, which support massive amount of randomized experiments run on data-driven enterprises. However, state-of-art approaches can become practically unreliable when the dimension of covariate increases to just 50, whereas experiments on large platforms can observe even  higher dimension of covariate. We propose a machine-learning-assisted covariate representation approach that can effectively make use of historical experiment or observational data that are run on the same platform to understand which lower dimensions can effectively represent the higher-dimensional covariate. We then propose design and estimation methods with the covariate representation. We prove statistically reliability and performance guarantees for the proposed methods. The empirical performance is demonstrated using numerical experiments.
}%


\KEYWORDS{Causal inference, machine learning, covariate representation} 
%

\maketitle

%


\section{Introduction}\label{sec:intro}
Utilizing  \textit{covariate} information can be a useful tool to improve the efficiency and accuracy for causal inference. The use of covariate for efficiency and accuracy improvement is prevalent and has been studied in both academic research (since \cite{rubin1974estimating}) and industry practices (\cite{deng2013improving,kohavi2020trustworthy}). Covariates are a group of information associated with each sample that may affect the distribution of outcomes or treatment assignments. For clinical trials, a sample represents a person and the covariate may include the person's age, pre-existing medical conditions and demographic information. For an e-commerce platform, a sample represents a customer or a visit to the website or mobile app, where the covariate may include the time of day, day of week, searched queries by a customer, etc.   
While the use of covariate has been shown to improve efficiency and accuracy, when the dimension of covariate becomes large, the effectiveness of utilizing covariate in experiment design and causal effect estimation encounter challenges. We describe two of such challenges as follows. 

\textbf{Challenge in experimental design for covariate balancing: } For experimental design, a randomization with significantly unbalanced groups on important covariates can lead to misleading conclusions. A popular method to improve the covariate balance is rerandomization using Mahalanobis distance (ReM, \cite{morgan2012rerandomization}), which discards treatment allocations with imbalanced covariates and only accepts those with covariates balanced enough. A quick recap of ReM is provided in Appendix \ref{append:intro}. Despite the importance of ReM, it becomes inefficient when the dimension of covariate grows, see Appendix \ref{append:intro} for numerical results and also \cite{wang2022rerandomization} for more theoretical evidence.

The fact is, the renowned covariate balancing methods can be ineffective when the covariate dimension starts to become moderately large. 

\textbf{Challenge in estimating causal effects: } Estimating causal effects is the core task in causal inference, where utilizing covariate information is critical not just for efficiency improvement but more importantly to improve accuracy and reduce confounding bias. The task includes estimating the average treatment effect (ATE), the conditional average treatment effect (CATE) and others. Dealing with higher-dimensional covariate can be challenging, and due to concerns about model misspecification, machine learning algorithms are commonly employed. This includes \cite{guo2017calibration,wager2018estimation,yao2018representation}, \cite{chernozhukov2018double},  \cite{hahn2020bayesian,kallus2020deepmatch}, and \cite{guo2021machine}. However, when high-dimensional inputs are involved in machine learning models, a large amount of data is required otherwise the performance of the learning algorithm could be poor.  While certain algorithms may offer asymptotic convergence results, achieving such results with high-dimensional covariates can often require too many data that are unrealistic. For example, the convergence result of causal forests (\cite{wager2018estimation}) requires ${\log (n)^d}/{n}\to 0$, where $n$ is the sample size and $d$ is the dimension of the covariates, and the required sample size $n$ can rarely be met in practice when $d>30$, a dimension that is not that large. 

If there is only one experiment with high-dimensional covariates and limited data samples, the aforementioned challenges are very difficult to solve. Thankfully, a large platform often runs more than tens of thousands of experiments every year and accumulates historical data from both experiments and daily operations. In fact, the collected historical data can often be relevant to a new experiment and contain information about how the high-dimensional covariate affects experiments. One rationale is that these data and experiments are collected and run on the same platform, which naturally share common structure and dependence. This observation motivates our efforts to leverage data from past experiments and observations to inform current experimental design and inference. 

In this work, we consider meta-learning via representation learning that aims to learn a shared low-dimensional representation of the covariates through past experiments and collected data. Our method is  related to meta-learning (\cite{finn2017model}), representation learning (\cite{bengio2013representation}), few-shot learning with representation (\cite{sun2017revisiting,goyal2019scaling,raghu2019rapid}), and low-dimensional representation learning (\cite{tripuraneni2020theory,du2020few,tripuraneni2021provable}), whereas our goal is to design experiments or estimate causal effects, different from them. There are some other methods to combine multiple collected data to estimate the treatment effects, see for example \cite{yang2020combining,athey2020combining}. These work make additional assumptions to find the relationship between outcomes of the collected data and the current task, while the only assumption we make is that there exists a generic covariate representation shared across all tasks.

\textbf{Our results}: (1) To assit experimental design and causal effect estimation, we introduce a machine learning assisted covariate representation approach that effectively utilizes historical experimental or observational data in similar scenarios. This approach helps identify lower-dimensional representations that can effectively capture the information present in higher-dimensional covariates. We then propose design and estimation methods that leverage the covariate representation. (2) We prove performance guarantees and provide statistical reliability for causal inference when using the covariate representation learned through our method. Our analysis proves that our approach can enhance efficiency in experimental design, improve accuracy in estimating Conditional Average Treatment Effects (CATE), and achieve near-optimality in estimating Average Treatment Effects (ATE). (3) 
Our message to industrial users is, our method could extract nonlinear representation of high-dimensional covariates from past experiments, which can enhance the performance of existing design and inference methods. 

\section{Problem Setup}\label{sec:setup}

In this section, we introduce a meta-learning framework aimed at learning a low-dimensional covariate representation from collected data, which often comprises high-dimensional covariates. Additionally, this framework yields a meta-model that offers enhanced efficiency when working with limited samples. To illustrate the fundamental concept and background, we provide the following examples.

\textbf{Example 1: (A/B tests in online platforms) } Many data-oriented companies conduct hundreds of A/B tests on their online platforms every day. Here, A/B tests refer to online experiments that randomly assign each customer to either a control (A) or treatment (B), collect outcomes, and conduct statistical inference on the average treatment effect. Although the outcomes of different A/B tests may have different distributions, it is common for these tests to be designed for a similar population (usually all customers), collect similar covariates (customer preferences, queries, etc), and use similar metrics (such as click-through rates and coversion rates). As a result, engineers have the opportunity to perform feature engineering to construct low-dimensional features from past experiments and expect these features to be helpful in future A/B tests. 


\textbf{Example 2: (Digital health experiments) } 
Increasing convenience and accuracy of wearable health-related devices have made it possible to run experiments digitally and more frequently than before, for example, to examine what approaches can better lower blood glucose after big meals. For these experiments, physical health information of patients is privately and locally collected and used as covariate information for analyzing the results. However, such covariate information can be high dimensional and pose challenges to accurately estimate the causal effects. 


The above two examples provide scenarios such that the data sets collected can have high-dimensional covariates that pose challenges to experiments and causal effect estimation. That said, these scenarios also represent cases where multiple historical data sets may be available, from which there is an opportunity to use meta learning with representation learning to extract the lower-dimensional covariate representation. In our meta learning framework, we do not assume that the outcomes share similar distribution across historical data sets. Instead, we only need the covariate information to share similar structure, which can be a more natural assumption, particularly for the aforementioned examples.




We now give the mathematical formulation of our problem. Suppose we are given a task set $\mathcal{T}=\{T_1,\cdots,T_K\}$ with $K$ tasks. For each $k=1,\cdots,K$, $T_k=\{(X_i^{(k)},I_i^{(k)},Y_i^{(k)})\}_{i=1}^{n_k}$ consists of $n_k$ observed triples. $X_i^{(k)}\in \mathcal{X}\subset\mathbb{R}^d$ is the set of covariates, $I_i^{(k)}\in\{0,1\}$ is the treatment indicator, and $Y_i^{(k)}\in\mathbb{R}$ is the outcome we observed. Here we assume all $X_i^{(k)}$ have the same dimension $d$, and a discussion for different dimensions is provided in Appendix \ref{append:different_d}. Though we assume $I\in\{0,1\}$ in this paper, our results can be naturally extended to the case with multiple treatments. We follow the potential outcome framework in \cite{rubin1974estimating} to write $Y_i^{(k)}=I_i^{(k)}Y_i^{(k)}(1)+(1-I_i^{(k)})Y_i^{(k)}(0)$, where $Y_i^{(k)}(1),Y_i^{(k)}(0)$ are potential outcomes of unit $i$ in the $k$-th task. We further assume that $\{(X_i^{(k)},I_i^{(k)},Y_i^{(k)}(1),Y_i^{(k)}(0))\}_{i=1}^{n_k}$ are independently and identically distributed (i.i.d.) according to some superpopulation probability measure $\mathbb{P}^{(k)}$, and all random variables in $T_k$ are independent of those in $T_j$ if $k\neq j$. We drop the subscript $i$ when discussing population stochastic properties of these quantities. Following two assumptions are standard in causal inference literature. 
\begin{assumption}\label{assump:unconfounded}
(Unconfoundedness). $(Y^{(k)}(1),Y^{(k)}(0))\indep I^{(k)}\vert X^{(k)}$ for $k=1,\cdots,K$.
\end{assumption}
\begin{assumption}\label{assump:overlap}
(Overlap). For $k=1,\cdots,K$, $0<P(I^{(k)}=1\vert X^{(k)})<1$ with probability 1.
\end{assumption}
We are interested in our current task $T_0$, and we have either $T_0=\{X_i^{(0)}\}_{i=1}^{n_k}$ for experimental design or $T_0=\{(X_i^{(0)},I_i^{(0)},Y_i^{(0)})\}_{i=1}^{n_0}$ for estimating treatment effects. We now introduce the high level idea of meta learning with covariate representation under this formulation. Rigorous assumptions and detailed theoretical results are deferred to Section \ref{sec:theory}. Our main assumption is that all tasks $T_k$ share the same covariate representation, that is, there exist a \textit{shared covariate representation} $\mathbf{h}:\mathbb{R}^d\to\mathbb{R}^r$ and $2(K+1)$ \textit{task-specific} functions $f_1^{(k)},f_0^{(k)}:\mathbb{R}^r\to\mathbb{R}$, such that 
$$\mathbb{E}^{(k)}[Y^{(k)}\vert X^{(k)}=x,I^{(k)}=l]= f_l^{(k)}\circ \mathbf{h}(x) \text{ for all }k=0,\cdots,K \text{ and } l=0,1.$$
Here $r\in\mathbb{N}_+$ is an integer much smaller than $d$. If we are not conducting experiments but analyzing observational data, then we would further assume that similar results hold for propensity scores, that is, there exist another  $\mathbf{g}:\mathbb{R}^d\to\mathbb{R}^{r'}$ and $K+1$ functions $e^{(k)}:\mathbb{R}^{r'}\to\mathbb{R}$, such that 
$
    \mathbb{P}^{(k)}[I^{(k)}=1\vert X^{(k)}=x]= e^{(k)}\circ \mathbf{g}(x)
$
and $r'<d$. Intuitively speaking, the expected outcome and the propensity score are affected by the covariate $x$ only through lower dimensional representations $(\mathbf{h}(x),\mathbf{g}(x))$. In Example 1, $\mathbf{h}(\cdot)$ is based on feature engineering and is usually fitted through some machine learning algorithm, and in Example 2 it is usually a variable selection function. If these assumptions are true, we could then learn an approximation $(\hat{\mathbf{h}},\hat{\mathbf{g}})$ of $(\mathbf{h},\mathbf{g})$, and $(\hat{\mathbf{h}},\hat{\mathbf{g}})$ can be applied to our task $T_0$.

\section{Our Algorithm}\label{sec:alg}

 We first discuss the procedure to learn the shared covariate representation $\mathbf{h}$ for the outcomes. For all tasks in $\mathcal{T}$, we consider the treatment group and the control group separately and split $T_k$ into $T_{k,1}=\{(X_i^{(k)},Y_i^{(k)})\vert 1\leq i\leq n_k,I_i^{(k)}=1\}$ and $T_{k,0}=\{(X_i^{(k)},Y_i^{(k)})\vert 1\leq i\leq n_k,I_i^{(k)}=0\}$. We then define $\mathcal{T}'=\{T_{k,l}\}_{k=1,\cdots,K; l=0,1}$. We use a function $\mathbf{h}_\theta: \mathbb{R}^d\rightarrow\mathbb{R}^s$, parameterized by $\theta$, to encode the covariates $X_i^{(k)}$ into a lower-dimensional representation $Z_i^{(k,l)}=\mathbf{h}_\theta(X_i^{(k)})$ for $k=1,\cdots,K$ and $I_i^{(k)}=l$. Here $h\in\mathcal{H}$ and $\mathcal{H}$ is by default chosen as the function class of shallow neural networks. Note that we do not know the dimension of real representation $r$ in advance, so here the output dimension $s$ is a hyperparameter. We suggest a conservative choice of $s$ to make sure $s\geq r$. In fact, if $\mathbf{h}:\mathbb{R}^d\to\mathbb{R}^r$ is the shared representation, then $\tilde{\mathbf{h}}(\cdot)=(\mathbf{h}(\cdot),\mathbf{0})$ with $\mathbf{0}\in\mathbb{R}^{s-r}$ is also a shared representation with dimension $s$. 
 
With the representation $Z_i^{(k,l)}$, we now train task-specific functions $f_{\phi_{k,l}}:\mathbb{R}^s\rightarrow\mathbb{R}$, parameterized by $\phi$, from a \textit{task-driven function class} $\mathcal{F}\subset C^1$, which maps $Z_i^{(k,l)}$ to the observed outcome $Y_i^{(k)}$. Here $C^1$ denotes the class of all continuously differentiable functions, and $\mathcal{F}$ is called ``task-driven" because it depends on how we use $\mathbf{h}_\theta(X_i^{(0)})$ in our task $T_0$, so it can be linear function class, parametric function class and neural networks. A detailed discussion on the choice of $\mathcal{F}$ is given in Appendix \ref{append:choice_F}. Now, aggregating the formulation, we make predictions on the outcome value in $T_{k,l}$ with covariate $X_i^{(k)}$ through $f_{\phi_{k,l}}\circ\mathbf{h}_\theta(X_i^{(k)})$. For a set of samples ${D}=\{(X_i^{(k)},Y_i^{(k)})\}$ randomly sampled from ${T}_{k,l}$, the loss $\mathcal{L}_{{T}_{k,l}}\left(f_{\phi_{k,l}}\circ\mathbf{h}_\theta,D\right)$ is defined with squared loss: 
\begin{equation}
    \mathcal{L}_{T_{k,l}}\left(f_{\phi_{k,l}}\circ\mathbf{h}_\theta,D\right) = \sum_{(X,Y)\in{D}}\big(Y-f_{\phi_{k,l}}\circ\mathbf{h}_\theta(X)\big)^2,
\end{equation}
and the total meta-learning loss is then given by
\begin{equation}
    \mathcal{L}_\text{Meta}(\mathcal{T}') = \mathbb{E}_{T_{k,l}\sim p(\mathcal{T}')}\mathcal{L}_{{T}_{k,l}}\left(f_{\phi_{k,l}}\circ\mathbf{h}_\theta,T_{k,l}\right).
\end{equation}
Here, $p(\mathcal{T}')$ is the task sampling distribution over $\mathcal{T}'$ and by default it is a uniform distribution over all $T_{k,l}$. In this case $\mathcal{L}_\text{Meta}(\mathcal{T}) = \frac{1}{2K}\sum_{k=1,\cdots,K;l=0,1}\mathcal{L}_{{T}_{k,l}}\left(f_{\phi_{k,l}}\circ\mathbf{h}_\theta,T_{k,l}\right)$ is simply the mean of losses of all tasks. With above notations, our algorithm is  given below in Algorithm \ref{alg:MAML}.

\begin{algorithm}[!htbp]
\caption{ MAML for Covariate Representation}
\label{alg:MAML}
\begin{algorithmic}[1]
\STATE{{Aim1: training general representation $\mathbf{h}_\theta$} (and the meta model $f_\phi$)}
\STATE{Require: $p(\mathcal{T}')$ : distribution over $\mathcal{T}'$, learning rates $\alpha_\text{in}, \alpha_\text{out}$, $\beta_\text{out}$} \\
\STATE{Randomly initialize $\theta$ and $\phi$}\\
\WHILE{not stop}
\STATE{Sample a batch of $b$ tasks $\{{T}^{(i)}\}_{i=1}^b \sim p(\mathcal{T}')$} \\
 \FORALL{${T}^{(i)}$} 
\STATE{Sample $K$ datapoints ${D}_i=\left\{({X}_j,{Y}_j)\right\}_{j=1}^K$ from ${T}^{(i)}$} \\
\STATE{Evaluate $\nabla_\phi \mathcal{L}_{\mathcal{T}^{(i)}}\left(f_\phi\circ \mathbf{h}_\theta,D_i\right)$} \\
\STATE{Update $\phi _i^{\prime}=\phi -\alpha_\text{in} \nabla_\phi \mathcal{L}_{\mathcal{T}^{(i)}}\left(f_\phi\circ \mathbf{h}_\theta,D_i\right)$}
\STATE{Sample datapoints ${D}_i^{\prime}=\left\{({X}_{j}', {Y}_{j}')\right\}$ from ${T}^{(i)}$ for the meta-update and representation-update} \\
\ENDFOR
\STATE{Update $\phi \leftarrow \phi-\alpha_\text{out} \nabla_\phi \sum_{i=1}^b \mathcal{L}_{{T}^{(i)}}\left(f_{\phi_i'}
\circ \mathbf{h}_\theta,D_i'\right)$} 
\STATE{Update $\theta \leftarrow \theta-\beta_\text{out} \nabla_\theta \sum_{i=1}^b \mathcal{L}_{{T}^{(i)}}\left(f_{\phi_i'}
\circ \mathbf{h}_\theta,D_i'\right)$}
\ENDWHILE
\STATE{Output: representation $\mathbf{h}_{{\theta}}$ and a meta model $f_{\phi}$}.
\STATE{\text{Aim2: adapting the meta model $f_\phi$ to task ${T}_{k,l} (0\leq k\leq K,l=0,1)$}}
\STATE{$\phi_{k,l}=\phi -\alpha_\text{in} \nabla_\theta \mathcal{L}_{{T}_{k,l}}\left(f_{\phi_*}\circ \mathbf{h}_\theta,T_{k,l}\right)$}
\end{algorithmic}
\end{algorithm}

 The meta-learning phase for the shared covariate representation and task-specific functions is closely relate to the Model-Agnostic Meta-Learning (MAML) algorithm proposed in \citep{finn2017model}. Given input datasets, our algorithm produces a representation function $\mathbf{h}_\theta$ as the main output, and also gives a meta model $f_\phi$ as a byproduct. The vector $\mathbf{h}_\theta$ can be employed to transform the covariates in our new tasks, generating low-dimensional representations that serve as replacements for the original covariates in downstream tasks. While the meta model $f_\phi$ is not indispensable for the downstream task, we can recover $f_{\phi_{k,l}}$ from it by taking gradient steps with samples in $T_{k,l},$ if necessary. Furthermore, the meta model can prove beneficial in predicting outcomes with limited samples, as demonstrated in Section \ref{sec:numerical}.
 In addition, while our focus is on randomized experiments, if $\mathcal{T}$ are data collected from observational studies and we need to estimate the propensity score in $T_0$, then we could use a procedure similar as Algorithm \ref{alg:MAML} to train a model which outputs a representation $\mathbf{g}_\beta$ and meta-model $e_{\psi}$ for the shared covariate representation $\mathbf{g}$ and task-specific functions $e^{(k)}$.

\section{Theoretical Results}\label{sec:theory}
 We prove theoretical guarantees for the the minimizer $\big((\hat{f}_{k,l}),\mathbf{h}_{\hat{\theta}}\big)$ of the meta loss $\mathcal{L}_\text{Meta}$, as a verification as statistical efficiency. Although a general convergence theory for meta learning is currently lacking, empirical success suggests that we can reasonably expect the output of Algorithm \ref{alg:MAML} to behave similarly to $\mathbf{h}_{\hat{\theta}}$. Thus, we will then focus on the results obtained using $\mathbf{h}_{\hat{\theta}}$ in causal inference tasks.

\subsection{Assumptions}
For simplicity, we assume $|T_{k,l}|=n$ for $k=1,\cdots,K$ and $l=0,1$. Also define $|T_{0}|=n_0$. To formalize the intuition of shared representation, we first adopt the following assumption, as presented in \cite{tripuraneni2020theory}.
\begin{assumption}\label{assump:representation}
(Shared representation). For all $k=0,\cdots,K$ and $l=0,1$, we have
\begin{equation}
    \mathbb{P}^{(k)}(X^{(k)}=x,Y^{(k)}=y\vert I^{(k)}=l)=\mathbb{P}_\mathcal{X}(x)\mathbb{P}_{Y\vert X,l}(y\vert r^{(k)}_l\circ\mathbf{h}^*(x)).
\end{equation}
Here $\mathbf{h}^* : \mathbb{R}^d \to \mathbb{R}^r$ is the shared covariate representation, and $r^{(k)}_l: \mathbb{R}^r\to \mathbb{R}$ are task-specific functions. $\mathbb{P}_\mathcal{X}$ is the marginal distribution of the covariate and $\mathbb{P}_{Y\vert X,l}$ is the conditional distribution of the outcome with $I=l$. 
\end{assumption}
Under Assumption \ref{assump:representation}, the conditional expectation $\mathbb{E}[Y^{(k)}\vert X^{(k)}=x,I^{(k)}=l]$ is a function of $r^{(k)}_l\circ\mathbf{h}^*(x)$, so we can write it as $f^{(k)}_l\circ\mathbf{h}^*(x)$ and we focus on $f^{(k)}_l$ instead of $r^{(k)}_l$ thereafter. We will also need following regularity conditions and realizability condition.
\begin{assumption}\label{assump:regular}
 (Regularity and realizability conditions). $Y^{(k)}$ are bounded for all $k=0,\cdots,K$; $\text{Cov}_{X\sim\mathbb{P}_\mathcal{X}}[\mathbf{h}^*(X)]$ is of full rank; For any $f\in \mathcal{F}$, $f$ is $L(\mathcal{F})$-Lipschitz with respect to ${l}_2$ norm; For any $f\in\mathcal{F},\mathbf{h}\in\mathcal{H}$, $f\circ \mathbf{h}$ is bounded over $\mathcal{X}$. In addition, the true representation $\mathbf{h}^*$ is contained in $\mathcal{H}$ and the true task-specific functions $f^{(k)}_l$ are contained in $\mathcal{F}$ for $k=0,\cdots,K$.
\end{assumption}
We will use the Gaussian complexity to measure the complexity of function classes $\mathcal{F,H}$. A quick recap of Gaussian complexity is provided in Appendix \ref{append:proofs}. Let $\mathcal{G}_n(\mathcal{Q})$ be the population Gaussian complexity of a function class $\mathcal{Q}$ with $n$ samples and
$\bar{\mathcal{G}}_n(\mathcal{Q})$ be the worst-case Gaussian complexity over $\mathcal{Q}$.
We now give the assumption on the complexity of $\mathcal{F}$ and $\mathcal{H}$.
\begin{assumption}\label{assump:complexity}
(Gaussian complexity bounds). 
    There exist $C_1,C_2>0$, $\gamma_1,\gamma_2\in (0,\frac{1}{2}]$ satisfying
    $$\bar{\mathcal{G}}_n(\mathcal{F})\leq \frac{C_1}{n^{\gamma_1}},\text{ and }\mathcal{G}_n(\mathcal{H})\leq \frac{C_2}{n^{\gamma_2}} \text{ for }n\in\mathbb{N}_+.$$
\end{assumption}
Note that for the majority of parametric function classes used in machine learning applications, Assumption \ref{assump:complexity} is satisfied by $\gamma_1=\gamma_2=\frac{1}{2}$ with $C$ be
the intrinsic complexity of the function classes. When $\mathcal{H}$ is the class of shallow neural networks, \cite{golowich2018size} shows that the assumption is also met with $\gamma_2=\frac{1}{2}$ under some additional conditions. Finally, we consider an assumption to capture the diversity of training tasks, and similar ideas are used in \cite{tripuraneni2020theory}.

\begin{assumption}\label{assump:task_diversity}
(Task diversity).
    For all $h'\in\mathcal{H},$ there exist $\nu_K,\epsilon_K>0$, such that
    $$\sup_{f_0\in\mathcal{F}}\inf_{f'\in\mathcal{F}}\mathbb{E}_{X\sim\mathbb{P}_\mathcal{X}}\big( f'\circ\mathbf{h}'(X)-f_0\circ\mathbf{h}^*(X)\big)^2\leq \frac{1}{2K\nu_K}\sum_{k,l}\inf_{f'\in\mathcal{F}}\mathbb{E}_{X\sim\mathbb{P}_\mathcal{X}}\big( f'\circ\mathbf{h}'(X)-f^{(k)}_l\circ\mathbf{h}^*(X)\big)^2+\epsilon_K.
    $$
\end{assumption}

\subsection{Causal inference with covariate representation}\label{subsec:CI_with_representation}
We now explore how to leverage the learned representation to perform causal inference tasks, such as experimental design and estimating treatment effects. All proofs are provided in Appendix \ref{append:proofs}. We first consider rerandomization using the Mahalanobis distance (ReM) in experimental design to balance the covariates, and we treat the learned representation as the covariates we want to balance. Since the effectiveness of ReM relies on the linear correlation between outcomes and covariates (\cite{li2018asymptotic}), we choose $\mathcal{F}$ to be the class of linear functions. 
The following result demonstrates that, in the case of a linear representation model, leveraging the learned representation allows for greater variance reduction.
\begin{theorem}\label{thm:ReM}
    Suppose $Y^{(k)}_i=\left(I^{(k)}_i\theta_{k1}+(1-I^{(k)}_i)\theta_{k0}\right)^{T}\mathbf{h}^*(X^{(k)}_i)$ follows a linear representation model, and Assumption \ref{assump:unconfounded}, \ref{assump:overlap} and \ref{assump:representation}$-$\ref{assump:task_diversity} hold. $\hat{\tau}$ is the difference-in-mean estimator of the treatment effect $\tau=\frac{\sum_{i=1}^{n_0}\big(Y^{(0)}_i(1)-Y^{(0)}_i(0)\big)}{n_0}$. Fix the task set $\mathcal{T}$ and $n$. Let the limiting regime be $n_0\to\infty$ and the proportion of treatment converges to $\rho\in(0,1)$. $\mathcal{D}_\text{Meta}(n),\mathcal{D}(n)$ are the asymptotic distributions of $\sqrt{n_0}(\hat{\tau}-\tau)$ under ReM with and without covariate representation. Then there exists some $C_3>0$, such that for any $\varepsilon>0$,
    \begin{equation}\label{eq:ReM}
        \lim_{n\to\infty}\mathbb{P}\left(\frac{{Var}(\mathcal{D}_\text{Meta}(n))}{{Var}(\mathcal{D}(n))}\leq\frac{F_{\chi^2_{s+2}}(F_{\chi^2_{s}}^{-1}(p))}{F_{\chi^2_{d+2}}(F_{\chi^2_{d}}^{-1}(p))}+C_3(\epsilon_K^{\frac{1}{2}}+\epsilon_K)+\varepsilon\right)=1,
    \end{equation}
    where $p$ is the acceptance probability in ReM, $s$ is the dimension of the covariate representation, $\chi^2_d$ is the chi-squared distrbution with $d$ degree of freedom, and its CDF is given by $F_{\chi^2_d}$.
\end{theorem}
With $d=500$, $s=20$, $p=0.001$, and by omitting the $\epsilon_K$ and $\varepsilon$ term, the RHS of the inequality in \eqref{eq:ReM} is approximately $0.33$, which means the asymptotic variance of ReM with covariate representation is one-third of that with original covariates in the case of a linear representation model. More numerical results for general models are provided in Section \ref{sec:numerical}.

We now consider estimating the CATE with learned representation $\mathbf{h}_{\hat{\theta}}$. The CATE given $x$ is
$\tau(x)=\mathbb{E}[Y^{(0)}(1)-Y^{(0)}(0)\big\vert X^{(0)}=x].$
Under Assumption \ref{assump:unconfounded}, we can estimate $\mathbb{E}[Y^{(0)}(l)\big\vert X^{(0)}=x]$ for $l=0,1$ separately. In this work we use following method that minimize the $l_2$ loss:
\begin{equation}\label{eq:fit}
\hat{f}^{(0)}_l=\argmin_{f\in\mathcal{F}}\sum_{(X,Y)\in T_{0,l}}\big(Y-f\circ\mathbf{h}_{\hat{\theta}}(X)\big)^2, \text{ for }l=0,1.
\end{equation}
We then estimate $\tau(x)$ through $\hat\tau(x)=\hat{f}^{(0)}_1\circ\mathbf{h}_{\hat{\theta}}(x)-\hat{f}^{(0)}_0\circ\mathbf{h}_{\hat{\theta}}(x)$, and we have:
\begin{theorem}\label{thm:CATE}
    Under Assumption \ref{assump:unconfounded}, \ref{assump:overlap} and \ref{assump:representation}$-$\ref{assump:task_diversity}, with probability $1-2\delta$ we have:
    \begin{equation}\label{eq:error_bound}
    \begin{aligned}
        \mathbb{E}(\hat{\tau}(X)-\tau(X))^2
        = O\left(\frac{\log(nK)}{\nu_K}\big(n^{-\gamma_1}+(nK)^{-\gamma2}\big)+n_0^{-\gamma_1}+\sqrt{\frac{\log(2/\delta)}{nK\nu_K^2}}+\sqrt{\frac{\log(2/\delta)}{n_0}}+\epsilon_K\right).\\
    \end{aligned}
    \end{equation}
    Here the expectation is taken over ${X\sim\mathbb{P}_\mathcal{X}}$ and all other randomness in $\mathcal{T}$ and $T_0$.
\end{theorem}
In most applications, the data we collect is often large, resulting in $n$ being significantly greater than $n_0$. As a result, the leading term in the RHS becomes $O(n_0^{-\gamma_1}+\sqrt{\frac{\log(2/\delta)}{n_0}}+\epsilon_K)$. Since in general $\gamma_1\leq\frac{1}{2}$, the error bound primarily consists of two components: the complexity of the learning problem with the function class $\mathcal{F}$, and the error arising from task diversity. The latter term serves as a replacement for the complexity of class $\mathcal{H}$.

Finally, we consider the problem of estimating the ATE in randomized experiments, where $\mathbb{P}(I=1\vert X)=p$ for some fixed $0<p<1$ and $I$ is independent of all other randomness. Discussions for general cases are provided in Appendix \ref{append:observational}. We estimate the average treatment effect $\tau=\mathbb{E}[Y(1)-Y(0)]$ through the widely used doubly-robust type estimator, which has the form as follows:
$$\hat{\tau}=\frac{1}{n_0}\sum_{(X,I,Y)\in T_{0}}\left(\frac{I(Y-\hat{Y}_1(X))}{\hat{p}}-\frac{(1-I)(Y-\hat{Y}_0(X))}{1-\hat{p}}+\hat{Y}_1(X)-\hat{Y}_0(X)\right),$$
here $\hat{Y}_l(X)$ is an estimator of $\mathbb{E}[Y(l)\vert X]$ and $\hat{p}=\sum_{i=1}^{n_0} I_i/n_0$. We assume $0<\hat{p}<1$ to make sure $\hat{\tau}$ is well-defined, which holds almost surely as $n_0\to\infty$. To construct $\hat{Y}_l(X)$, we follow the idea of \cite{chernozhukov2018double} to divide $T_{0,l}$ into $M$ folds, denoted as $T_{0,l}=\cup_{i=1}^MS_{i,l}$, here $M$ is a hyperparameter and is commonly chosen to be $5$. For each $(X,Y)\in S_{i,l}$, we fit $\hat{f}^{(0)}_{l,i}$ on $T_{0,l}\setminus S_{i,l}$ using a similar approach as shown in \eqref{eq:fit} and construct $\hat{Y}_l(X)=\hat{f}^{(0)}_{l,i}\circ \mathbf{h}_{\hat{\theta}}(X)$.
While the doubly-robust estimator is known to be unbiased and, under mild conditions, asymptotically normally distributed even with model misspecification, we are interested in quantifying the amount of variance reduction achieved through our meta-learning approach for covariate representation. This is proved by following results.
\begin{theorem}\label{thm:ATE}
Under Assumption \ref{assump:unconfounded}, \ref{assump:overlap} and \ref{assump:representation}$-$\ref{assump:task_diversity}, with the task set $\mathcal{T}$ and $n$ fixed, we have $\sqrt{n_0}(\hat{\tau}-\tau)=\mathcal{D}_{n_0}+\mathcal{E}_{n_0}$. Here
$$\mathcal{D}_{n_0}\stackrel{d}\rightarrow \mathcal{N}(0,V_\text{optimal}) \text{  as }n_0\to\infty,$$
and $V_\text{optimal}=\mathbb{E}_{X\sim \mathbb{P}_\mathcal{X}}\big[\frac{Var[Y(1)\vert X]}{p}+\frac{Var[Y(0)\vert X]}{1-p}+(\tau(X)-\tau)^2\big]$ is the semiparametric lower bound of the asymptotic variance. The term $\mathcal{E}_{n_0}$ represents the error with the property $\mathbb{E}[\mathcal{E}_{n_0}]=0$, and $\mathbb{E}[\mathcal{E}^2_{n_0}]$ is bounded by the same term as in \eqref{eq:error_bound} with probability $1-2\delta$.
\end{theorem}
Theorem \ref{thm:ATE} demonstrates that with the learned representation, $\hat{\tau}$ is nearly optimal, as its asymptotic variance reaches the semiparametric lower bound, with the exception of an error introduced by the meta-learning process. The error diminishes as both $n$ and $n_0$ grow larger, and as the tasks in $\mathcal{T}$ become more diverse.

\section{Numerical Results}\label{sec:numerical}
We present numerical results on simulated data in this section. Additional numerical results are provided in Appendix \ref{append:experiments}.
\subsection{Simulation}
\subsubsection{Rerandomization using Mahalanobis distance (ReM) for general models}\label{sec: simulation ReM}
In the simulation experiment, we test different underlying covariate representations $\mathbf{h}^*$, including using all covariates, variable selection among all covariates, linear combination of covariates, and representation mapping through a neural network. The task-specific function $f$ is a logistic-type function. The sample features have an original dimension of $d=300$ and the underlying representation dimension is $r=50$. With all mentioned above, the outcome $Y^{(k)}$ is generated through $Y^{(k)}=\frac{1}{1+\exp{ \left(a^{(k)\top} \mathbf{h}^*(X)+b^{(k)}\right)}}+\epsilon $, where $\epsilon\sim \mathcal{N}(0,0.01)$, and the parameters are sampled independently through $a^{(k)}\sim (U(-1,1))^{300}, b^{(k)} \sim U(-1,1)$. 
During the meta-learning phase, we generated $20$ tasks, each containing $1000$ samples and their corresponding outcomes. We use these tasks to learn the representation and evaluate ReM approach on an additional dataset.

To learn an approximation of the representation $\mathbf{h}^*$, we employed a three-hidden-layer fully connected neural networks (FCN) with ReLU activation function. The covariates were encoded into a representation with dimension $s$ for all tasks. Additionally, we used a two-hidden-layer neural network with Tanh activation to approximate the task-specific function ${f}^{(k)}$. 
\begin{table}[htbp]
    \centering
    \begin{tabular}{c|c|c|c}
    \hline
    & Original    &  Representation ($s=50$) &   Representation ($s=30$) \\ \hline
Full variables   & 0.788  &  0.677  & 0.746 \\ \hline
Variable selection &   0.801  & 0.706  & 0.779\\  \hline
Linear combination &  0.813   &  0.737  & 0.756 \\  \hline
Neural network &  0.884  &  0.746  &  0.767 \\  \hline
    \end{tabular}
    \caption{The ratio of variance under ReM with different underlying covariate representation}
    \label{table:rerandomization simulation}
\end{table}
Table \ref{table:rerandomization simulation} demonstrates that utilizing and balancing the learned representation  can substantially reduce variance, compared to standard use of covariate information, irrespective of the underlying ground-truth generating feature. The meta-learning algorithm, when combined with a learned representation, effectively captures nonlinearity and extracts a lower-dimensional representation from meta tasks.
\subsubsection{Estimating the conditional average treatment effect}
\begin{figure}[htbp]
\centering
\begin{minipage}[t]
{0.48\textwidth}
\centering
\includegraphics[width=6cm]{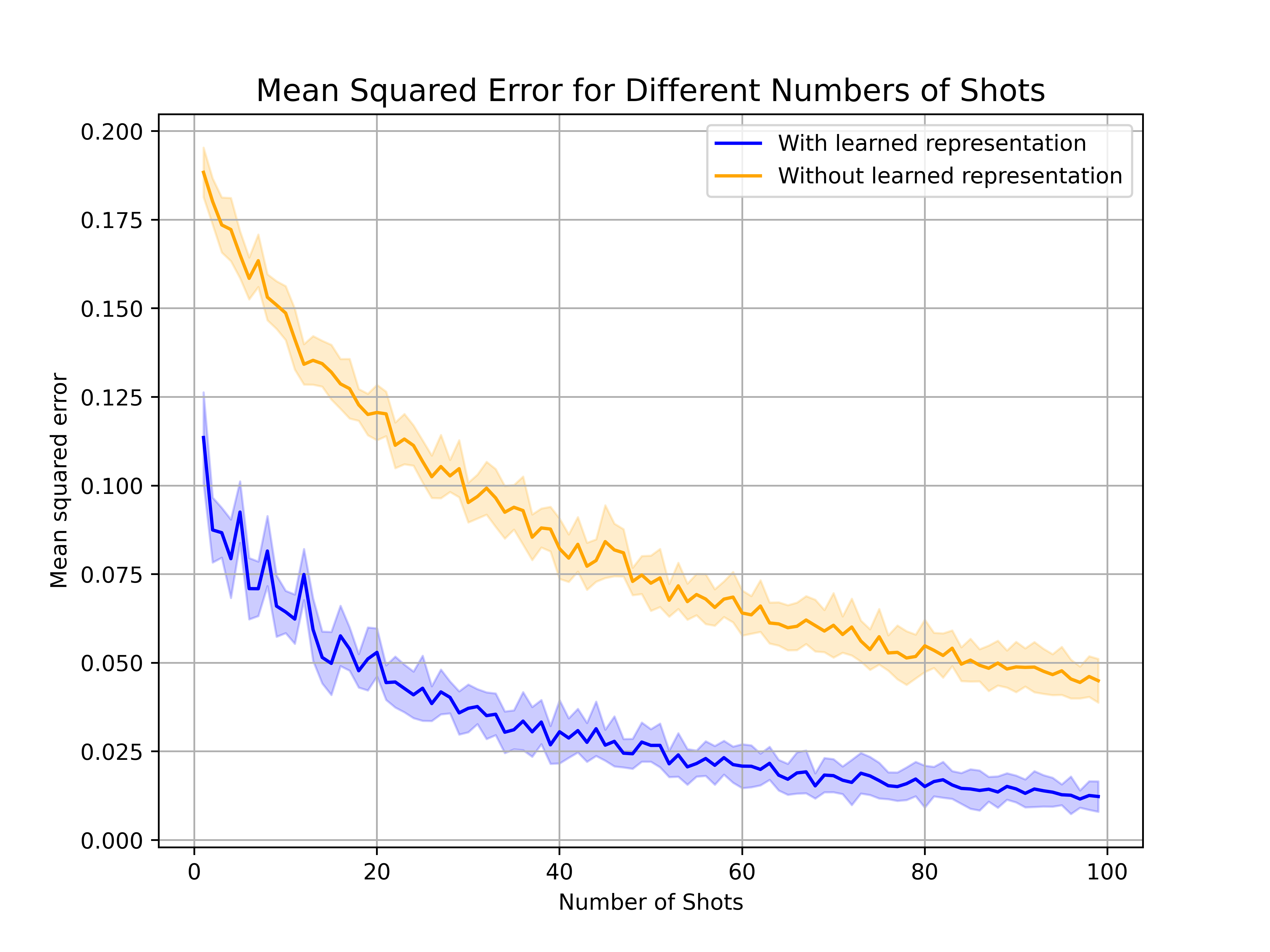}
\end{minipage}
\begin{minipage}[t]{0.48\textwidth}
\centering
\includegraphics[width=6cm]{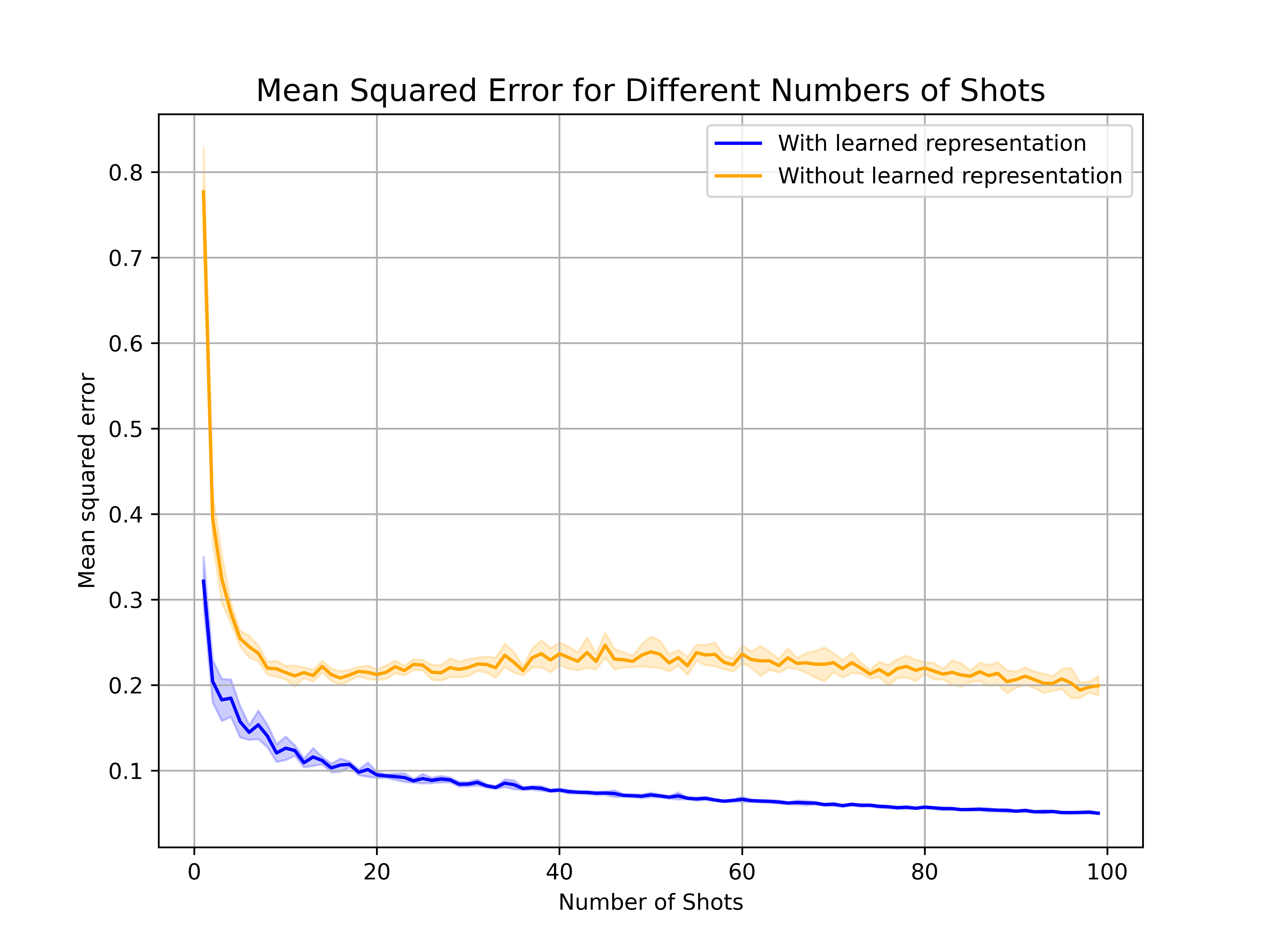}
\end{minipage}
\caption{MSE of CATE; \\Left: linear representation; Right: ANN representation}\label{fig: CATE 2 simulation}
\end{figure}

In this section, we consider the task of estimating the Conditional Average Treatment Effect (CATE) as described in Section \ref{sec:theory}. We consider a total of $K=20$ tasks, each comprising $1000$ samples. 
The outcome model, covariate representation function and task-specific functions follow the same structure as described in Section \ref{sec: simulation ReM}. For validation purposes, we incorporate an additional validation task and record the MSE of the CATE estimator on the validation set using different sample sizes when adapting the task-specific function. Figure \ref{fig: CATE 2 simulation} presents the results, considering two different underlying feature representations: a linear mapping and a neural network. Both mappings transform the original covariates from $\mathbb{R}^{300}$ to $\mathbb{R}^{50}$. In both case, our method achieves much smaller MSE compared with baseline method that does not use a learned representation.

\vspace{-2mm}

\section{Conclusion}
We introduce a machine learning assisted covariate representation approach in order to address challenges in causal inference with high-dimensional covariate. We prove statistical reliability and performance guarantees of the proposed methods. We validate the efficacy of our method on numerical simulations.

\bibliographystyle{plainnat}
\bibliography{arXiv_version}

\newpage

\section*{Supplementary Material}

\section{Supplementary materials for Section \ref{sec:intro}}\label{append:intro}
{\noindent\bf A quick recap of rerandomization with Mahalanobis distance (ReM)}

Consider a randomized experiment with $m$ units, with $m_1$ assigned to treatment and $m_0$ assigned to control, $m=m_1+m_0$. Before the experiment, we collect covariates $X_i=(X_{1i},\cdots,X_{si})^T$ for the $i$-th units. $I=(I_1,\cdots,I_m)$ is the treatment assignment vector, where $I_i=1$ means unit $i$ is assigned to the treatment, and $I_i=0$ indicates a control. Define $\hat{\tau}_X=\frac{1}{m_1}\sum_{i=1}^m I_iX_i-\frac{1}{m_0}\sum_{i=1}^m (1-I_i)X_i$. We consider Rubin's potential outcome model and write the potential outcome of unit $i$ under treatment and control as $Y_i(1)$ and $Y_i(0)$, respectively. The individual treatment effect is defined as $\tau_i=Y_i(1)-Y_i(0)$, and the average treatment effect of $m$ units is $\tau=\frac{1}{m}\sum_{i=1}^m\tau_i$ and is what we are interested in. We now define $S_X^2=\frac{1}{m-1}\sum_{i=1}^m(X_i-\bar{X})(X_i-\bar{X})^T$ to be the sample covariance matrix of $X$, here $\bar{X}=\frac{1}{m}\sum_{i=1}^mX_i$. Let $r_1=\frac{m_1}{m}$ and $r_0=\frac{m_0}{m}$, $V_{xx}$ is defined as $V_{xx}=\frac{S_X^2}{r_1r_0}$. We can now describe the procedure of ReM. We define the following Mahalanobis distance between the
covariate means in treatment and control groups:
\begin{equation}
    M=(\sqrt{m}\hat{\tau}_X)^TV_{xx}^{-1}(\sqrt{m}\hat{\tau}_X).
\end{equation}
A treatment assignment $I$ is accepted only if $M\leq a$ for some prespecified threshold $a$. In this way, ReM accepts only those randomizations with the Mahalanobis distance less than or equal to $a$, so the treatment assignments with significantly imbalanced covariates will be discarded.

{\noindent\bf Variance reduction of rerandomization with different dimensions}

\begin{figure}[ht]
\centering
\caption{The percent reduction in variance when $R^2=0.5$ and $p=0.01$}
\includegraphics[scale=0.7]{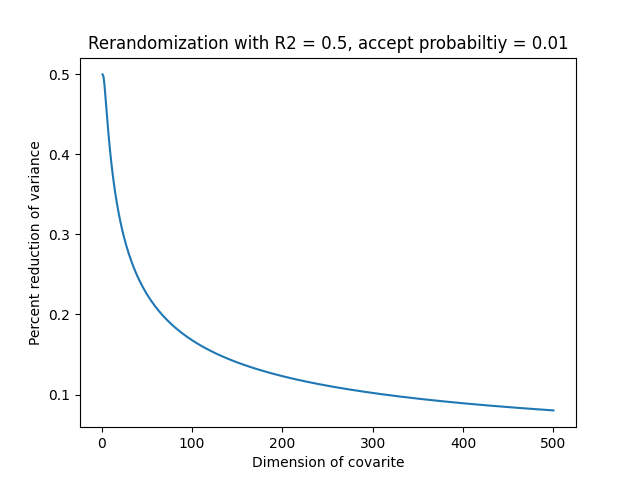}
\label{fig:rerandomization_vr}
\end{figure}

\begin{figure}[ht]
\centering
\caption{The percent reduction in variance when $R^2=0.5$ and $p=0.001$}
\includegraphics[scale=0.7]{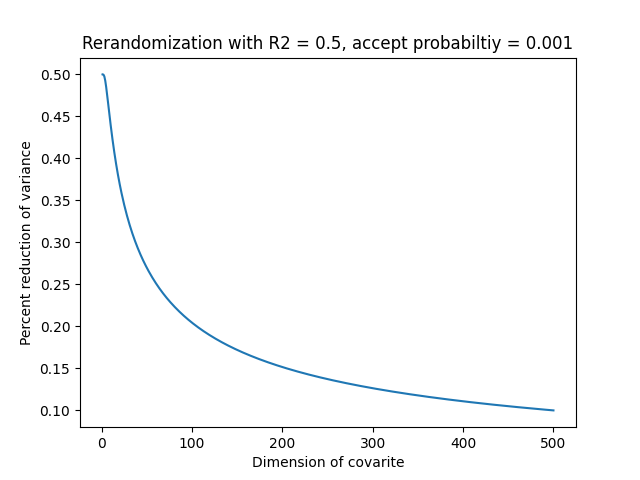}
\label{fig:rerandomization_vr_05_e-3}
\end{figure}

\begin{figure}[ht]
\centering
\caption{The percent reduction in variance when $R^2=0.2$ and $p=0.001$}
\includegraphics[scale=0.7]{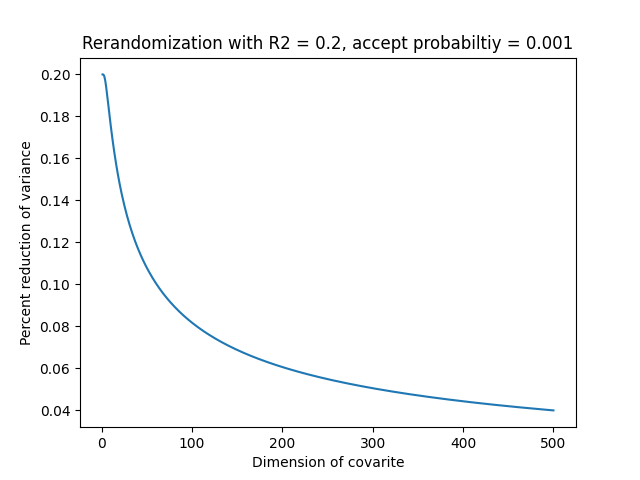}
\label{fig:rerandomization_vr_02_e-3}
\end{figure}

Figure \ref{fig:rerandomization_vr},\ref{fig:rerandomization_vr_05_e-3} and \ref{fig:rerandomization_vr_02_e-3} show the percent reduction in variance of ReM with different $R^2$ and acceptance proability $p$. We can see that whatever $R^2$ and $p$ is, the percent reduction goes to $0$ very fast as the dimension of the covariates grows. 

\section{Supplementary materials for Section \ref{sec:setup}}
{\noindent\bf Discussions on tasks with different covariates}\label{append:different_d}

In real-world applications, the covariates can vary across tasks. This implies that the dimension of covariates and the specific meaning of each dimension differ for $X$ in different tasks. To address this issue, our algorithm pads the original covariate $\left\{X^{(k)}\right\} \in \mathbb{R}^{d_k}$ in the $k^\text{th}$ task with $d_\text{max}-d_k$ \textbf{missing covariates} into a higher-dimensional covariate $\tilde{X}^{(k)} \in \mathbb{R}^{d_\text{max}}$, where the $d_\text{max} - d_k$ dimensions are padded from covariates in other tasks. The padded $\tilde{X}^{(k)} \in \mathbb{R}^{d_\text{max}}$ has the same dimension $d_\text{max}$ for all tasks, with each dimension representing the same feature. Consequently, the original task $T_k$ is transformed into $\tilde{T}_k = {(\tilde{X}_i^{(k)}, I_i^{(k)}, Y_i^{(k)}, M_i^{(k)})}_{i=1}^{n_k}$, where $M_i^{(k)}\in\mathbb{R}^{d_\text{max}}$ indicates which covariates are missing in task $k$. For the missing covariates in task $k$, we fill them with either $0$ or the average value from all other covariates in the tasks with the corresponding features. Subsequently, the training process for representation follows the same steps as outlined in Algorithm \ref{alg:MAML}, with the replacement of $X_i^{(k)}$ with $\tilde{X}_i^{(k)}$.

We examine the padding approach through simulations using the same underlying outcome and covariate generating functions introduced in Section \ref{sec: simulation ReM}. There are a total of $400$ possible covariates. For each task $k$, we generate $\tilde{X}^{(k)}\in (U[-1,1])^{400}\subset \mathbb{R}^{400}$, where $U[-1,1]$ denotes a uniform distribution between $-1$ and $1$. The underlying covariate representation is generated using the four methods proposed in Section \ref{sec: simulation ReM}, and the corresponding outcome $Y^{(k)}$ is generated from a random logistic function.

Randomly selecting $d_k\in[100,300]$ covariates, we construct $X^{(k)}$ along with $Y^{(k)}$ to form each task $k$. In the simulation experiment, we generate $20$ tasks, each consisting of $1000$ samples. An additional task is generated for evaluation purposes, and the results of the ReM in the evaluation task are reported in Table \ref{table:ReM for missing}.

\begin{table}[htbp]
    \centering
    \begin{tabular}{c|c|c|c}
    \hline
    & Original    &  Representation ($s=80$) &   Representation ($s=40$) \\ \hline
Full variables   & 0.824  &  0.742  & 0.766 \\ \hline
Variable selection &   0.855  & 0.762  & 0.809\\  \hline
Linear combination &  0.864   &  0.796  & 0.786 \\  \hline
Neural network &  0.902  &  0.825  &  0.847 \\  \hline
    \end{tabular}
    \caption{The ratio of variance under ReM with missing covariates.}
    \label{table:ReM for missing}
\end{table}

\section{Supplementary materials for Section \ref{sec:alg}}
{\noindent\bf The choice of $\mathcal{F}$ in the meta-learning phase}\label{append:choice_F}

We discuss the heuristic choice of task-driven function class $\mathcal{F}$. Note that after the representation $\mathbf{h}_\theta$ is learned, we would then apply it to our current task $T_0$ through replacing $X_i^{(0)}$ by $\mathbf{h}_\theta(X_i^{(0)})$. $\mathcal{F}$ is called ``task-driven" because it depends on how we use $\mathbf{h}_\theta(X_i^{(0)})$ in our task $T_0$. Suppose $T_0=\{X_i^{(0)}\}_{i=1}^{N_0}$ and we want to balance the covariates in $T_0$ with rerandomization, then we would take $\mathcal{F}$ to be the class of linear functions, since rerandomization is robust to misspecification and its variance reduction gain depends on the linear correlation of outcomes and covariates (\cite{li2018asymptotic}). This naturally leads us to search for the best representation $\mathbf{h}_\theta(X)$ for the linear regression model. By taking $\mathcal{F}$ to be the class of linear functions, our meta-learning algorithm aims to find the best linear approximations for all training tasks. We thus expect that the learned representation will perform well on the downstream rerandomization task. The power of this choice is illustrated through some numerical experiments in Section \ref{sec:numerical}. Similar choice also holds for randomized experiments with linear adjustment (\cite{rosenbaum2002covariance,deng2013improving}), which also aims to utilize the linear correlation of outcomes and covariates. 

In many applications, like estimating the uplift (CATE) \cite{gutierrez2017causal} or using double machine leanring estimator \cite{chernozhukov2018double} to estimate the treatment effect, people may want to use the representation as the input of some learning algorithms to predict the outcome, then taking $\mathcal{F}$ to be the class of functions the same as the function class used in the downstream learning task would be a good choice, as long as it is differentiable. It can commonly be chosen as a specific class of parametric functions, or even a class of neural networks.

\section{Supplementary materials for Section \ref{sec:theory}}\label{append:proofs}
{\noindent\bf A quick recap of Gaussian complexity}

For a generic vector-valued function class $\mathcal{Q}$ consist of functions $\mathbf{q}:\mathbb{R}^{r_1}\to\mathbb{R}^{r_2}$ and $N$ data points $\mathbf{X}=(X_1,\cdots,X_N)^T$, then the empirical Gaussian complexity is defined as:
$\hat{\mathcal{G}}_{\mathbf{X}}(\mathcal{Q})=\mathbb{E}\big[\sup_{\mathbf{q}\in\mathcal{Q}}\sum_{k=1}^{r_2}\sum_{i=1}^Nq_k(X_i)\varepsilon_{k,i}\big],$
where $\varepsilon_{k,i}\stackrel{\text{i.i.d.}}\sim \mathcal{N}(0,1)$ and $\mathbf{q}_k$ is the $k$-th coordinate of $\mathbf{q}$. We define the corresponding population Gaussian complexity as $\mathcal{G}_N(\mathcal{Q})=\mathbb{E}_\mathbf{X}[\hat{\mathcal{G}}_{\mathbf{X}}(\mathcal{Q})]$. We further define the worst-case Gaussian complexity over $\mathcal{F}$ as:
$$\bar{\mathcal{G}}_n(\mathcal{F})=\max_{\mathbf{Z}\in \mathcal{Z}}\hat{\mathcal{G}}_{\mathbf{X}}(\mathcal{F}), \text{ where } \mathcal{Z}=\{\mathbf{h}(X_1),\cdots,\mathbf{h}(X_n)\vert X_i\in \mathcal{X},h\in\mathcal{H}\}.$$

{\noindent\bf Proof of Theorem \ref{thm:ReM}}

Through out this proof, we write the learned representation $\mathbf{h}_{\hat\theta}$ as $\mathbf{h}$ to simplify the notation. Suppose we assign $m_1$ samples to treatment and $m_0$ to control with $m_1+m_0=n_0$. We consider Rubin's potential outcome model and write the potential outcome of unit $i$ under treatment and control as $Y_i(1)$ and $Y_i(0)$, respectively. The individual treatment effect is defined as $\tau_i=Y_i(1)-Y_i(0)$, and the average treatment effect of $n_0$ units is $\tau=\frac{1}{n_0}\sum_{i=1}^{n_0}\tau_i$. Define $S_{Y(l)}^2=\sum_{i=1}^{n_0}\frac{\left(Y_i(l))-\bar{Y}(l)\right)^2}{n_0-1}$
to be the sample variance of $Y(l)$ for $l=0,1$. Also define $S_{\mathbf{h}(X)}^2=\frac{1}{n_0}\sum_{i=1}^{n_0}(\mathbf{h}(X_i)-\bar{\mathbf{h}}(X))(\mathbf{h}(X_i)-\bar{\mathbf{h}}(X))^T$, $S_{\tau}^2=\frac{1}{n_0-1}\sum_{i=1}^{n_0}(\tau_i-\tau)^2$, $S_{Y(l),\mathbf{h}(X)}=S_{\mathbf{h}(X),Y(l)}^T=\frac{1}{n_0-1}\sum_{i=1}^{n_0}(Y_i(l)-\bar{Y}(l))(\mathbf{X_i}-\bar{\mathbf{h}}(X))^T$. $r_1=\frac{m_1}{n_0}$ and $r_0=\frac{m_0}{n_0}$. $S_{Y(l)\vert \mathbf{h}(X)}^2=S_{Y(l),\mathbf{h}(X)}(S_{\mathbf{h}(X)}^2)^{-1}S_{\mathbf{h}(X),Y(l)}$ and similarly define $S^2_{\tau\vert X}$. The dimension of $\mathbf{h}(X)$ is $s$, and the acceptance threshold in ReM is taken to be $a_s$ such that the acceptance probability is $p$. 

Given above definitions, under ReM with learned representation, the asymptotic distribution of $\sqrt{n_0}(\hat{\tau}-\tau)$ under ReM is given by Theorem 1 of \cite{li2018asymptotic} as:
\begin{equation}\label{eq:proof_ReM_clt}
    \sqrt{n_0}(\hat{\tau}-\tau)\stackrel{d}\to \sqrt{V_{\tau\tau}}\left(\sqrt{1-R^2}\varepsilon_0+\sqrt{R^2}L_{s,a_s}\right),
\end{equation}
where $V_{\tau\tau}=\frac{S_{Y(1)}^2}{r_1}+\frac{S_{Y(0)}^2}{r_0}-S_{\tau}^2$, $\varepsilon_0\sim\mathcal{N}(0,1)$ and is independent of $L_{s,a}$, $L_{s,a}\sim E_1\vert E^TE\leq a$ with $E=(E_1,\cdots,E_s)^T\sim \mathcal{N}(0,\mathbf{I}_s)$, and
\begin{equation}\label{eq:proof_ReM_R2}
    R^2=\frac{r_1^{-1}S_{Y(1)\vert \mathbf{h}(X)}^2+r_0^{-1}S_{Y(0)\vert \mathbf{h}(X)}^2-S_{\tau\vert \mathbf{h}(X)}^2}{r_1^{-1}S_{Y(1)}^2+r_0^{-1}S_{Y(0)}^2-S_{\tau}^2}.
\end{equation}
Since $$
\begin{aligned}
    S_{Y(1)\vert \mathbf{h}(X)}^2&=S_{Y(1),\mathbf{h}(X)}(S_{\mathbf{h}(X)}^2)^{-1}S_{\mathbf{h}(X),Y(1)},\\
\end{aligned}$$
by LLN we have
$$
\begin{aligned}
    S_{Y(1)\vert \mathbf{h}(X)}^2\stackrel{a.s.}\rightarrow A\stackrel{\text{def}}=Cov(Y(1),\mathbf{h}(X))Var(\mathbf{h}(X))^{-1}Cov(\mathbf{h}(X),Y(1))^T\\
\end{aligned}$$
as $n_0\to\infty$. Since $Y(1)=\theta_{01}^T\mathbf{h}^*(X)$, we can write it as $Y(1)=\theta_{01}^T\mathbf{h}^*(X)=\hat{\theta}_{01}^T\mathbf{h}(X)+\Delta(X)$, where $\hat{\theta}^T\mathbf{h}(X)$ is the best linear predictor that minimize the $l_2$ loss with representation $\mathbf{h}(X)$. Then $Cov(Y(1),\mathbf{h}(X))=\hat{\theta}_{01}^TVar(\mathbf{h}(X))+Cov(\Delta(X),\mathbf{h}(X)),$
so $|A-Var(\hat{\theta}_{01}^T\mathbf{h}(X))|\leq C_0\left(\mathbb{E}[\Delta(X)+\Delta^2(X)]\right)$. Similarly, we have $S^2_{Y(1)}\stackrel{a.s.}\rightarrow Var(Y(1))=Var(\hat{\theta}_{01}^T\mathbf{h}(X)+\Delta(X))$,
and $\vert Var(\hat{\theta}_{01}^T\mathbf{h}(X)+\Delta(X))-Var(\hat{\theta}_{01}^T\mathbf{h}(X))\vert\leq C_1\left(\mathbb{E}^{\frac{1}{2}}[\Delta^2(X)]+\mathbb{E}[\Delta^2(X)]\right).$
Thus, 
$$1-\frac{S_{Y(1)\vert \mathbf{h}(X)}^2}{S^2_{Y(1)}}\stackrel{a.s.}\rightarrow \Delta_A\leq C_2\left(\mathbb{E}^{\frac{1}{2}}[\Delta^2(X)]+\mathbb{E}[\Delta^2(X)]\right),$$
for some $C_2>0$.
The above arguments also hold for ${S_{Y(0)\vert \mathbf{h}(X)}^2}$ and ${S_{Y(1)\vert \tau}^2}$, plug all these results into \eqref{eq:proof_ReM_R2} and we have 
$$R^2\stackrel{a.s.}\rightarrow1+\Delta_{R^2},$$
where $\Delta_{R^2}\leq C_4\left(\mathbb{E}^{\frac{1}{2}}[\Delta^2(X)]+\mathbb{E}[\Delta^2(X)]\right)$.
Now we have
\begin{equation}\label{eq:proof_ReM_3}
    \frac{{Var}(\mathcal{D}_\text{Meta}(n))}{{Var}(\mathcal{D}(n))}\leq \frac{V_{\tau\tau}\left(1-R^2(1-Var(L_{s,a_s}))\right)}{V_{\tau\tau}(Var(L_{d,a_d}))}\leq \frac{Var(L_{s,a_s})}{Var(L_{d,a_d})}+C_5\left(\mathbb{E}^{\frac{1}{2}}[\Delta^2(X)]+\mathbb{E}[\Delta^2(X)]\right).
\end{equation}
\cite{li2017general} and Proposition 2 of \cite{li2018asymptotic} show that the first term in the RHS of \eqref{eq:proof_ReM_3} is 
given by $\frac{F_{\chi^2_{s+2}}(F_{\chi^2_{s}}^{-1}(p))}{F_{\chi^2_{d+2}}(F_{\chi^2_{d}}^{-1}(p))}$. As for the second term, we use the results of Theorem 3 in \cite{tripuraneni2020theory}. Note that their assumptions are met under our assumptions, then with probability $1-2\delta$,
\begin{equation}
    \mathbb{E}[\Delta^2(X)]\leq O\left(\frac{\log(nK)}{\nu_K}\big(n^{-\gamma_1}+(nK)^{-\gamma2}\big)+n_0^{-\gamma_1}+\sqrt{\frac{\log(2/\delta)}{nK\nu_K^2}}+\sqrt{\frac{\log(2/\delta)}{n_0}}+\epsilon_K\right).
\end{equation}
Since in our limiting regime we take $n_0\to\infty$, and by then taking $n\to\infty$ we know the second term becomes $O(\epsilon_K^{\frac{1}{2}}+\epsilon_K)+o_p(1)$. Plug into \eqref{eq:proof_ReM_3} and we obtain what we want.

{\noindent\bf Proof of Theorem \ref{thm:CATE}}

We use the results of Theorem 3 in \cite{tripuraneni2020theory}. Note that their assumptions are met under our Assumption \ref{assump:representation}, \ref{assump:regular} and \ref{assump:task_diversity}. Then they show that, with probability $1-2\delta$, the risk for the current task $T_{0,l}$ with $l\in\{0,1\}$ is upper bounded by 
\begin{equation}\label{eq:proof_CATE_1}
    \begin{aligned}
        &\mathbb{E}\left(Y-\hat{f}_l^{(0)}\circ\mathbf{h}_{\hat{\theta}}(X)\right)^2-\mathbb{E}\left(Y-{f}_l^{(0)}\circ\mathbf{h}^*(X)\right)^2\\
        &\leq 
        O\left(\frac{\log(nK)}{\nu_K}\big(\bar{\mathcal{G}}_n(\mathcal{F})+{\mathcal{G}}_{nK}(\mathcal{H})\big)+\bar{\mathcal{G}}_{n_0}(\mathcal{F})+
        \frac{1}{\nu_K(nK)^2}+
        \sqrt{\frac{\log(2/\delta)}{nK\nu_K^2}}+\sqrt{\frac{\log(2/\delta)}{n_0}}+\epsilon_K\right).
    \end{aligned}
\end{equation}   
Here the expectation is taken over $X\sim\mathbb{P}_{\mathcal{X}}$ and all randomness in $\mathcal{T}$ and $T_0$, and by Assumption \ref{assump:regular} we know $\mathbf{h}^*\in \mathcal{H}$.

Under Assumption \ref{assump:complexity}, we have $\bar{\mathcal{G}}_n(\mathcal{F})=O(n^{-\gamma_1}),\bar{\mathcal{G}}_{n_0}(\mathcal{F})=O({n_0}^{-\gamma_1})$ and $\mathcal{G}_{nK}(\mathcal{H})=O((nK)^{-\gamma_2})$, plug into \eqref{eq:proof_CATE_1} and omit the higher order term $\frac{1}{\nu_K(nK)^2}$ we obtain that
\begin{equation}
    \begin{aligned}
 &\mathbb{E}\left(Y-\hat{f}_l^{(0)}\circ\mathbf{h}_{\hat{\theta}}(X)\right)^2-\mathbb{E}\left(Y-{f}_l^{(0)}\circ\mathbf{h}^*(X)\right)^2\\
        &\leq 
        O\left(\frac{\log(nK)}{\nu_K}\big(n^{-\gamma_1}+(nK)^{-\gamma2}\big)+n_0^{-\gamma_1}+\sqrt{\frac{\log(2/\delta)}{nK\nu_K^2}}+\sqrt{\frac{\log(2/\delta)}{n_0}}+\epsilon_K\right).
    \end{aligned}
\end{equation}
Now, since $\mathbb{E}[Y(l)\vert X]=f_l^{(0)}\circ \mathbf{h}^*(X)$, by the bias-variance decomposition we have
\begin{equation}\label{eq:proof_CATE_2}
    \mathbb{E}\left(Y-\hat{f}_l^{(0)}\circ\mathbf{h}_{\hat{\theta}}(X)\right)^2-\mathbb{E}\left(Y-{f}_l^{(0)}\circ\mathbf{h}^*(X)\right)^2=\mathbb{E}\left(\hat{f}_l^{(0)}\circ\mathbf{h}_{\hat{\theta}}(X)-{f}_l^{(0)}\circ\mathbf{h}^*(X)\right)^2.
\end{equation}
Plug \eqref{eq:proof_CATE_2} into \eqref{eq:proof_CATE_1} we obtain:
\begin{equation}\label{eq:proof_CATE_3}
    \begin{aligned}
&\mathbb{E}\left(\hat{f}_l^{(0)}\circ\mathbf{h}_{\hat{\theta}}(X)-{f}_l^{(0)}\circ\mathbf{h}^*(X)\right)^2\\
        &\leq 
        O\left(\frac{\log(nK)}{\nu_K}\big(n^{-\gamma_1}+(nK)^{-\gamma2}\big)+n_0^{-\gamma_1}+\sqrt{\frac{\log(2/\delta)}{nK\nu_K^2}}+\sqrt{\frac{\log(2/\delta)}{n_0}}+\epsilon_K\right).
    \end{aligned}
\end{equation}
Now, we have
\begin{equation}\label{eq:proof_CATE_4}
\begin{aligned}
    &\mathbb{E}\left(\hat{\tau}(X)-\tau(X)\right)^2\\    &=\mathbb{E}\left(\left(\hat{f}_1^{(0)}\circ\mathbf{h}_{\hat{\theta}}(X)-{f}_1^{(0)}\circ\mathbf{h}^*(X)\right)-\left(\hat{f}_0^{(0)}\circ\mathbf{h}_{\hat{\theta}}(X)-{f}_0^{(0)}\circ\mathbf{h}^*(X)\right)\right)^2\\
    &\leq 2\mathbb{E}\left(\hat{f}_1^{(0)}\circ\mathbf{h}_{\hat{\theta}}(X)-{f}_1^{(0)}\circ\mathbf{h}^*(X)\right)^2+2\mathbb{E}\left(\hat{f}_0^{(0)}\circ\mathbf{h}_{\hat{\theta}}(X)-{f}_0^{(0)}\circ\mathbf{h}^*(X)\right)^2.
\end{aligned}
\end{equation}
Now, we can use \eqref{eq:proof_CATE_3} to bound the RHS of \eqref{eq:proof_CATE_4} and we obtain what we want.

{\noindent\bf Proof of Theorem \ref{thm:ATE}}

We begin by write $\hat{\tau}$ as $\hat{\tau}=\tilde{\tau}+\delta_1-\delta_0$, where
\begin{equation}
    \tilde{\tau}=\frac{1}{n_0}\sum_{(X,I,Y)\in T_{0}}\left(\frac{I(Y-\mathbb{E}[{Y}(1)\vert X])}{\hat{p}}-\frac{(1-I)(Y-\mathbb{E}[{Y}(0)\vert X])}{1-\hat{p}}+\mathbb{E}[{Y}_1(X)\vert X]-\mathbb{E}[{Y}_0(X)\vert X]\right),
\end{equation}
\begin{equation}
    \delta_1 = \frac{1}{n_0}\sum_{(X,I,Y)\in T_0}\left(1-\frac{I}{\hat{p}}\right)\left(\hat{Y}_1(X)-\mathbb{E}[Y(1)\vert X]\right),
\end{equation}
and \begin{equation}
    \delta_0 = \frac{1}{n_0}\sum_{(X,I,Y)\in T_0}\left(1-\frac{1-I}{1-\hat{p}}\right)\left(\hat{Y}_0(X)-\mathbb{E}[Y(0)\vert X]\right).
\end{equation}
We take $\mathcal{D}_{n_0}=\sqrt{n_0}(\tilde{\tau}-\tau)$ and $\mathcal{E}_{n_0}=\sqrt{n_0}(\delta_1-\delta_0)$. For $\mathcal{E}_{n_0}$, note that
$$\mathbb{E}\left[1-\frac{I}{\hat{p}}\right]=\mathbb{E}\left[\mathbb{E}\left[1-\frac{I}{\hat{p}}\Big\vert \hat{p}\right]\right]=0,$$
so $\mathbb{E}[\mathcal{E}_{n_0}]=\mathbb{E}[\delta_1-\delta_0]=0$. In addition, we have
\begin{equation}\label{eq:proof_ATE_1}
    \mathbb{E}[\delta_1^2\vert \hat{p}]=\frac{1-\hat{p}}{n_0\hat{p}}\mathbb{E}\left[\hat{Y}_1(X)-\mathbb{E}[Y(1)\vert X]\right]^2+\frac{n_0-1}{n_0}\mathbb{E}\left[(1-\frac{I}{\hat{p}})(1-\frac{I'}{\hat{p}})\Big\vert \hat{p}\right]\left(\mathbb{E}[\hat{Y}_1(X)]-\mathbb{E}[Y(1)]\right)^2.
\end{equation}
Direct calculation gives $$\mathbb{E}\left[(1-\frac{I}{\hat{p}})(1-\frac{I'}{\hat{p}})\Big\vert \hat{p}\right]=-1+\frac{(n_0\hat{p}-1)}{(n_0-1)\hat{p}}=\frac{\hat{p}-1}{(n_0-1)\hat{p}},$$
plug into \eqref{eq:proof_ATE_1} and we have:
$$n_0\mathbb{E}[\delta_1^2\vert \hat{p}]=\frac{1-\hat{p}}{\hat{p}}\mathbb{E}\left[\hat{Y}_1(X)-\mathbb{E}[Y(1)\vert X]\right]^2-\frac{1-\hat{p}}{\hat{p}}\left(\mathbb{E}[\hat{Y}_1(X)]-\mathbb{E}[Y(1)]\right)^2.$$
Since $\mathbb{P}(|\hat{p}-p|>\varepsilon)$ is exponentially small, there exists some constant $C_1>0$ such that
$$n_0\mathbb{E}[\delta_1^2]\leq C_1\mathbb{E}\left[\hat{Y}_1(X)-\mathbb{E}[Y(1)\vert X]\right]^2.$$
Similarly, we have 
$$n_0\mathbb{E}[\delta_0^2]\leq C_2\mathbb{E}\left[\hat{Y}_0(X)-\mathbb{E}[Y(0)\vert X]\right]^2.$$
Combine above results together we then obtain:
\begin{equation}\label{eq:proof_ATE_2}
    \mathbb{E}[\mathcal{E}_{n_0}^2]\leq C_3\left(\mathbb{E}\left[\hat{Y}_1(X)-\mathbb{E}[Y(1)\vert X]\right]^2+\mathbb{E}\left[\hat{Y}_0(X)-\mathbb{E}[Y(0)\vert X]\right]^2\right).
\end{equation}
Note that $\hat{Y}_l(X)$ is given by $\hat{f}_{l,i}^{(0)}\circ \mathbf{h}_{\hat{\theta}}(X)$, and the data used to fit $\hat{f}_{l,i}^{(0)}$ is independent of $X$, so the same bound in \eqref{eq:proof_CATE_3} also applies to \eqref{eq:proof_ATE_2}, so we obtain the same bound for $\mathbb{E}[\mathcal{E}_{n_0}^2]$ as in \eqref{eq:error_bound} with probability $1-2\delta$.

We now consider the $\mathcal{D}_{n_0}$ term. We first decompose $\tilde{\tau}$ as $\tilde{\tau} = \tau^*+e$, where
\begin{equation}
    \tau^*=\frac{1}{n_0}\sum_{(X,I,Y)\in T_{0}}\left(\frac{I(Y-\mathbb{E}[{Y}(1)\vert X])}{p}-\frac{(1-I)(Y-\mathbb{E}[{Y}(0)\vert X])}{1-p}+\mathbb{E}[{Y}_1(X)\vert X]-\mathbb{E}[{Y}_0(X)\vert X]\right),
\end{equation}
and 
\begin{equation}
    e = \frac{p-\hat{p}}{n_0\hat{p}p}\sum\big(I(Y-\mathbb{E}[Y(1)])\big)-\frac{p-\hat{p}}{n_0(1-\hat{p})(1-p)}\sum\big((1-I)(Y-\mathbb{E}[Y(0)])\big).
\end{equation}
Since $p-\hat{p}=O_p(n_0^{-\frac{1}{2}})$, $I(Y-\mathbb{E}[Y(1)])$ are i.i.d, and $(1-I)(Y-\mathbb{E}[Y(0)])$ is i.i.d., we know $e=o_p(n^{-\frac{1}{2}})$, so the limiting distribution of $\mathcal{D}$ is determined by the limiting distribution of $\sqrt{n_0}(\tau^*-\tau)$. Note that now each term in the summation of $\tau^*$ is i.i.d., by central limit theorem for i.i.d. random variables we have:
\begin{equation}
    \sqrt{n_0}(\tau^*-\tau)\stackrel{d}\rightarrow \mathcal{N}(0,V_\text{optimal}),
\end{equation}
where $V_\text{optimal}=\mathbb{E}_{X\sim \mathbb{P}_\mathcal{X}}\big[\frac{Var[Y(1)\vert X]}{p}+\frac{Var[Y(0)\vert X]}{1-p}+(\tau(X)-\tau)^2\big]$ is known as the semiparametric lower bound of the ATE estimator, see for example \cite{hahn1998role}. This finishes our proof.

{\noindent\bf Estimating causal effects for general scenarios}\label{append:observational}

In general cases, we can still utilize the doubly-robust estimator to estimate the ATE. However, we cannot expect our estimator to be unbiased since the empirical estimator $\hat{p}$ for $p(I|X)$ is no longer valid. In this case, we need to replace it with the estimated propensity score.
To address this, we can follow a similar approach as we did for the outcome model. First, we learn a low-dimensional representation from the collected data using meta-learning. Then, we estimate the propensity score using the learned representation. If similar assumptions hold for our propensity score estimator, it should closely approximate the true propensity score, resulting in a good ATE estimator.
We illustrate the performance of this method through additional numerical experiments in Appendix \ref{append:experiments}. The method works well across different underlying representations, and in all scenarios, the MSE of our estimator is significantly smaller than that of estimators without using a learned representation. However, the theoretical analysis of our estimator in general scenarios is left as future work.

\section{Supplementary Experiments}\label{append:experiments}
\subsection{Simulation}
\subsubsection{Estimating average treatment effect in general scenarios}

In this section, we examine the Average Treatment Effect (ATE) estimator in two scenarios: (a) randomized experiments with a fixed treatment assignment probability $p$, and (b) observational studies with propensity score $p^{(k)}(x)$. For both scenarios, we construct the doubly-robust estimator proposed in Section \ref{sec:theory}.
In problem (a), we utilize the empirical mean for $\hat{p}$. In problem (b), we train an additional neural network based on the learned representation $\hat{p}^{(k)}(\mathbf{h}_\theta(X))$ to approximate the propensity score for given covariates $X$ in task $k$. 

\begin{figure}[htbp]
\centering
\begin{minipage}[t]
{0.48\textwidth}
\centering
\includegraphics[width=6cm]{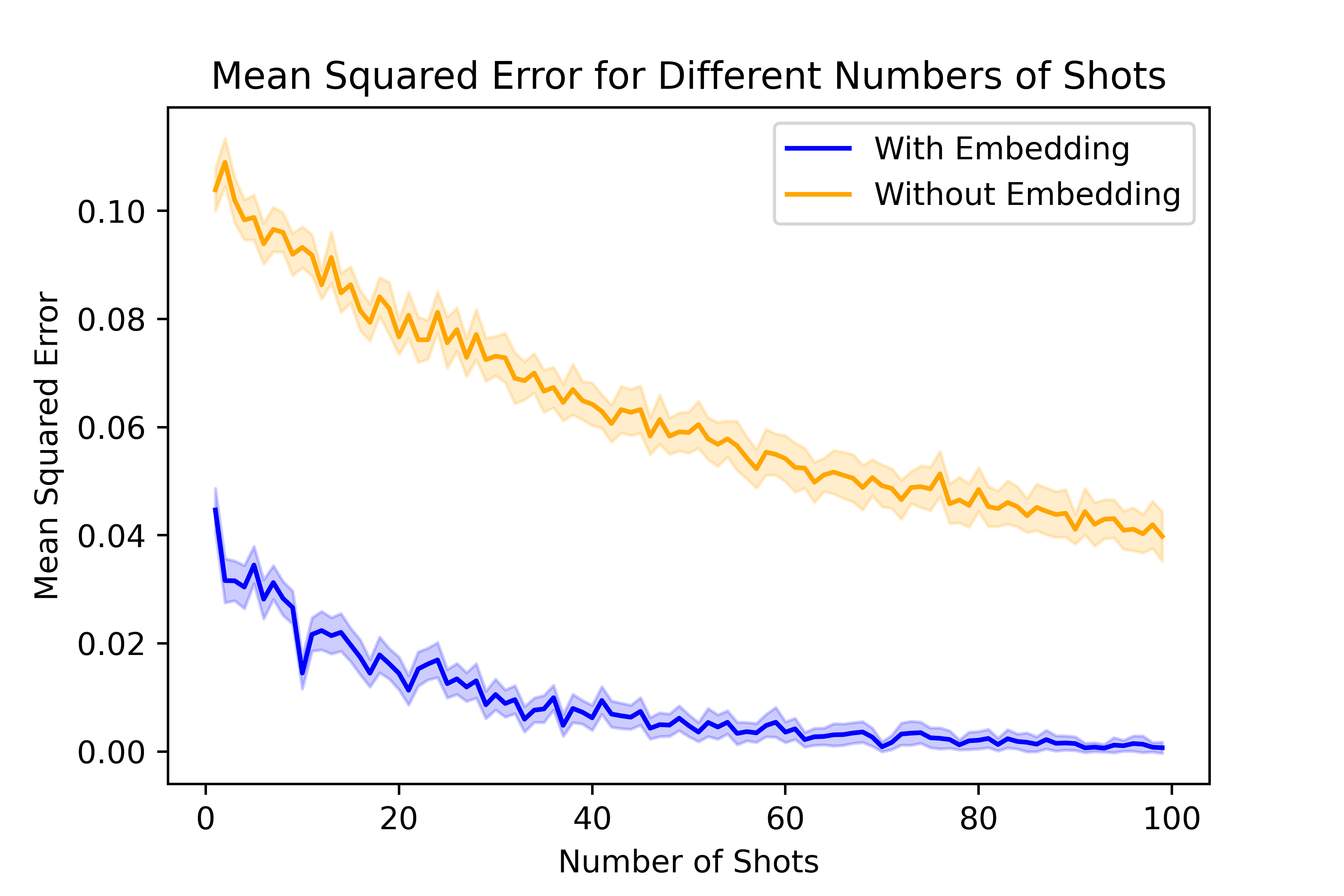}
\end{minipage}
\begin{minipage}[t]{0.48\textwidth}
\centering
\includegraphics[width=6cm]{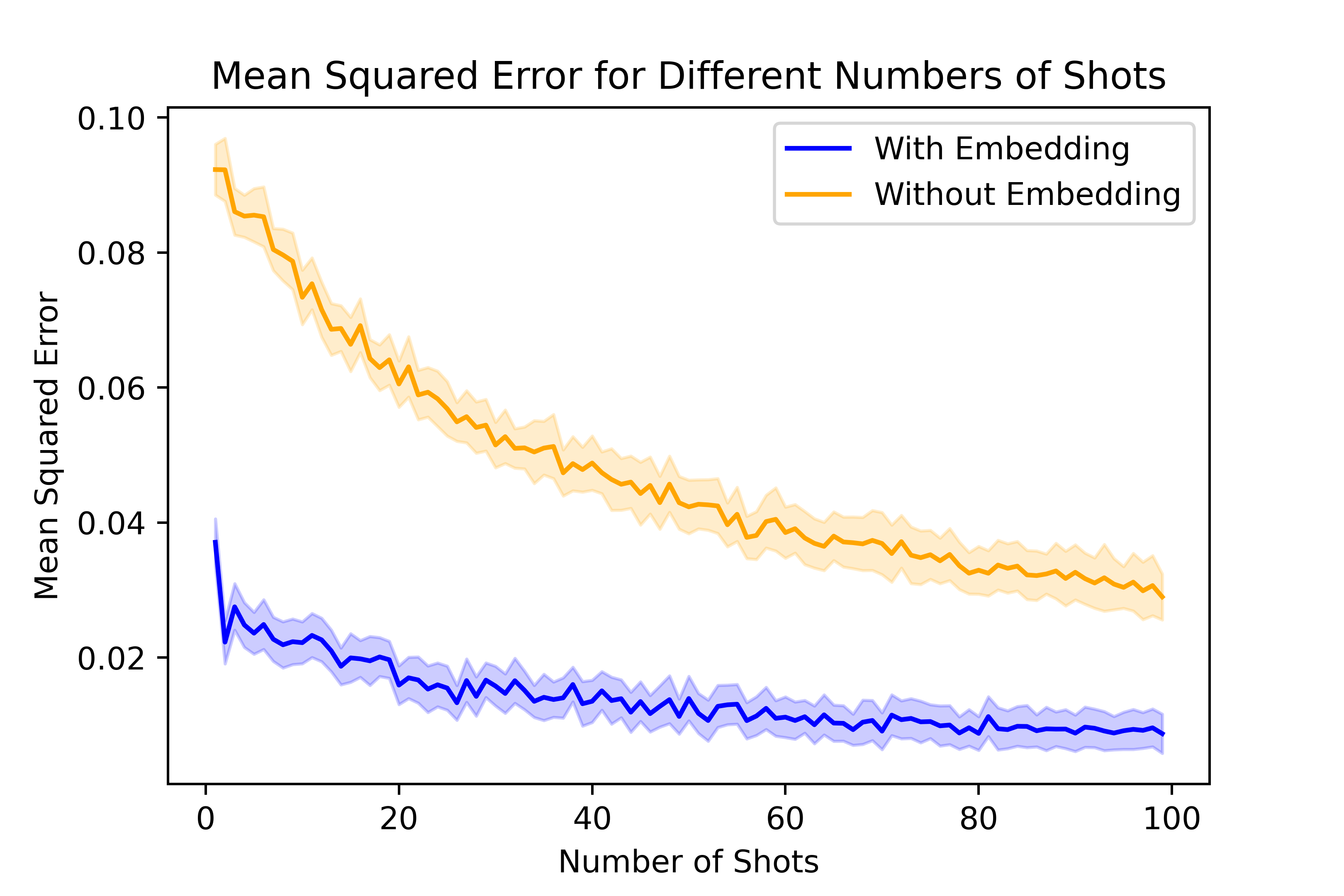}
\end{minipage}
\caption{MSE of ATE \\
Left: underlying representation as 100 selected features;\\ Right: underlying representation as 30 selected features}\label{fig: append fixed p ATE}
\end{figure}

\begin{figure}[htbp]
\centering
\begin{minipage}[t]
{0.48\textwidth}
\centering
\includegraphics[width=6cm]{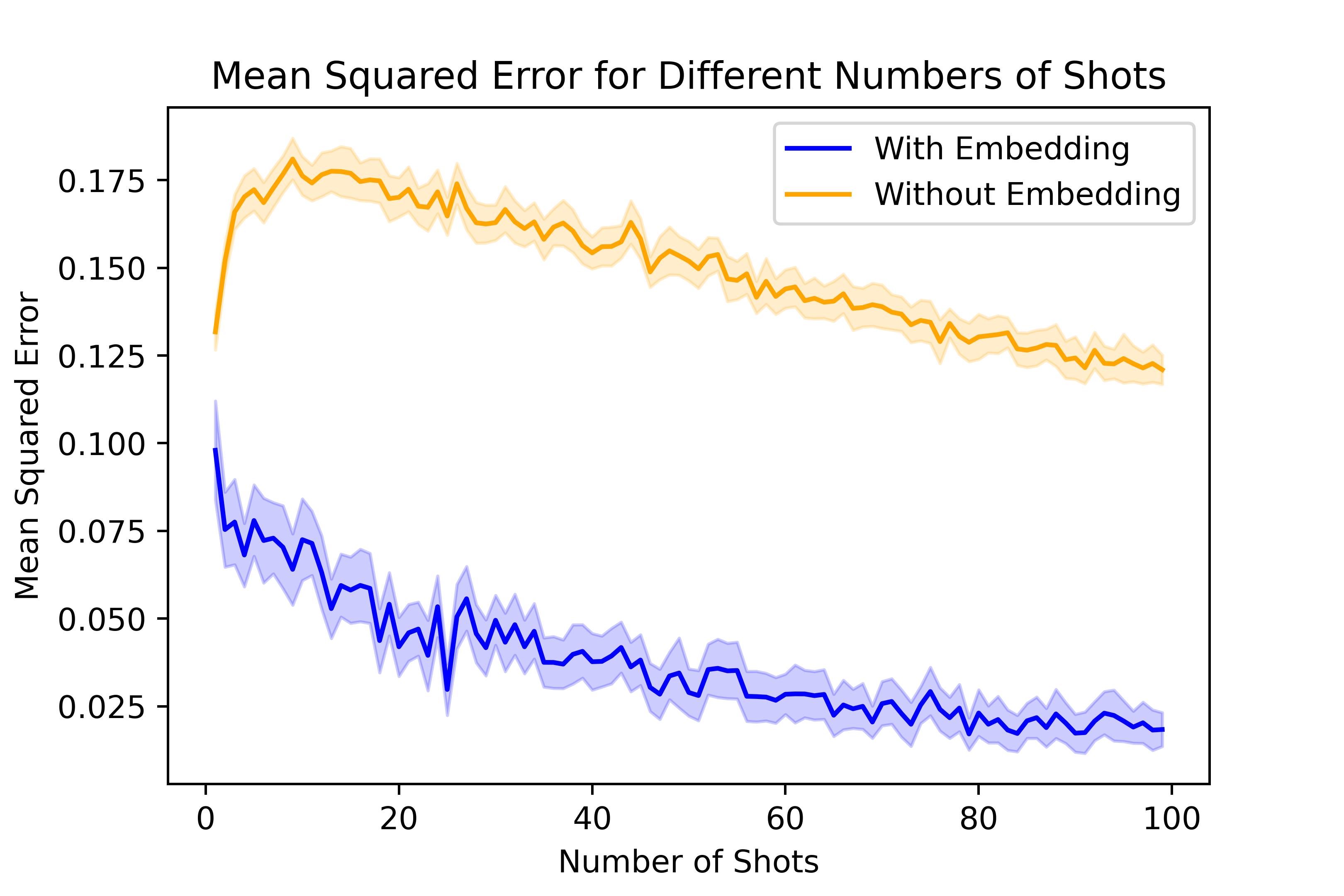}
\end{minipage}
\begin{minipage}[t]{0.48\textwidth}
\centering
\includegraphics[width=6cm]{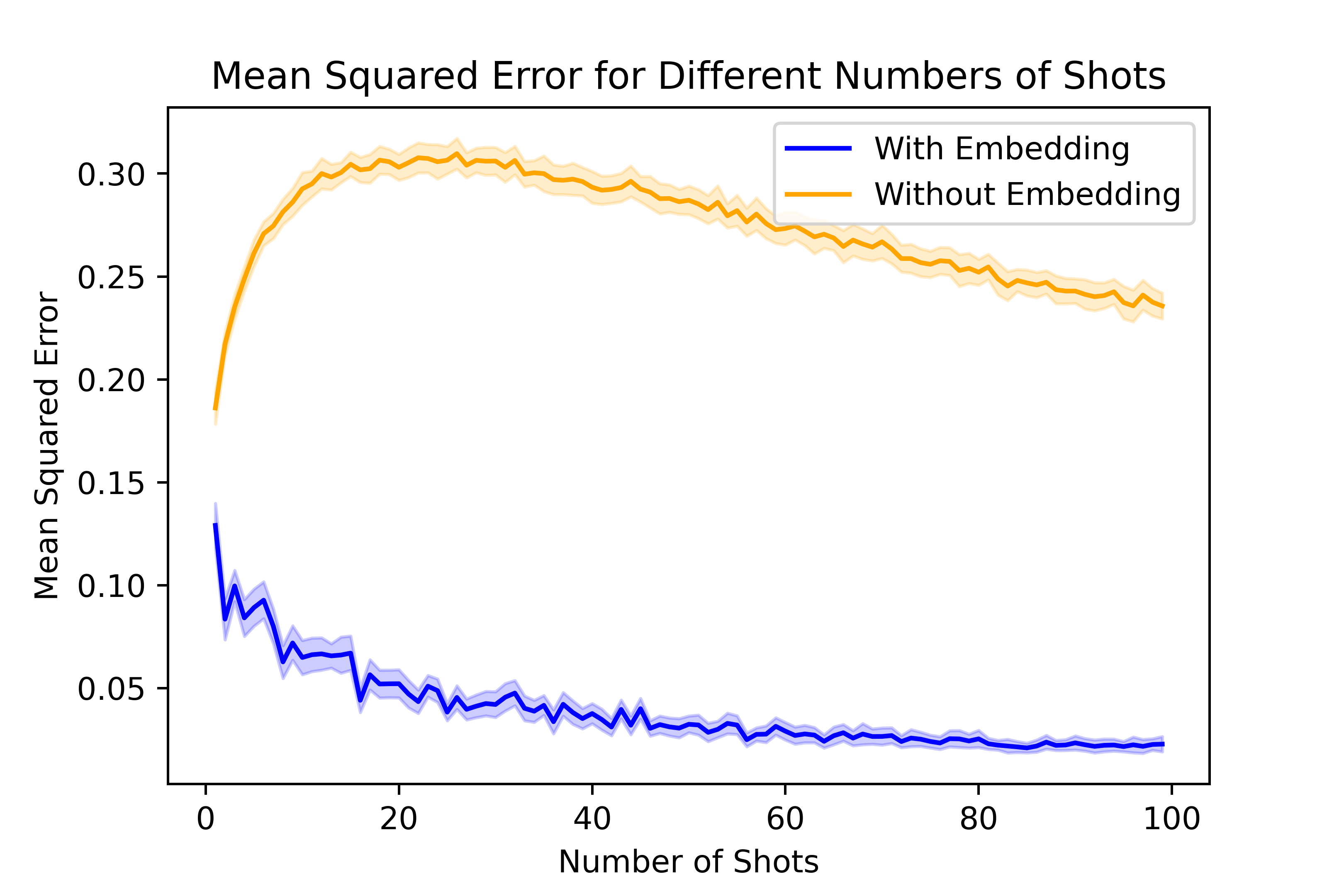}
\end{minipage}
\caption{MSE of ATE \\
Left: linear representation; Right: ANN representation}\label{fig: append with propensity function ATE}
\end{figure}

For problem (a), we randomly select $p \in [0.2,0.8]$ as the ground-truth treatment probability. In problem (b), the propensity score is generated from a random neural network $p(x)$, where the output is constrained between $[0,1]$ for any covariates $X$. The outcomes for the control group $Y^{(0)}$ and the $k^\text{th}$ treatment (task $k$) $Y^{(k)}$ are generated from the function $f$ introduced in Section \ref{sec: simulation ReM}. In the simulation for (a) and (b), we generate $K=40$ tasks, each with a different $p$ (or $p(X)$). We evaluate the Mean Squared Error (MSE) of the ATE estimator under different samples (shots) in an additional validation task. For baseline algorithm without feature representation, we directly trained two neural networks with input as original covariates $X$ to separately estimate the outcome function and the propensity socre. We repeat the experiments 10 times and present the average MSE together with 95\% interval error-bar in Figure \ref{fig: append fixed p ATE} and \ref{fig: append with propensity function ATE}. 

Our approach generally obtain better performance in estimating ATE in a new task, especially when the number of samples is small. For problem (b), the baseline method suffers from `increasing errors' under small-sample settings, which may be induced by the fact that we need to estimate the propensity function and outcome function simultaneously. This requires a larger sample size for training. Our approach mitigates this issue by providing a better pre-trained feature representation.

\vspace{-2mm}

\end{document}